\documentclass[showpacs,preprintnumbers,amsmath,amssymb,prb,twocolumn]{revtex4}

\def\XXint#1#2#3{{\setbox0=\hbox{$#1{#2#3}{\int}$}
     \vcenter{\hbox{$#2#3$}}\kern-.5\wd0}}

\usepackage{graphicx}
\usepackage{subfigure}
\usepackage{bm}

\begin{document}

\title{Cubic B20 helimagnets with quenched disorder in magnetic field}

\author{O.\ I.\ Utesov$^{1,2}$}
\email{utiosov@gmail.com}
\author{A.\ V.\ Syromyatnikov$^{1,3}$}
\email{asyromyatnikov@yandex.ru}

\affiliation{$^1$National Research Center ``Kurchatov Institute'' B.P.\ Konstantinov Petersburg Nuclear Physics Institute, Gatchina 188300, Russia}
\affiliation{$^2$St. Petersburg Academic University - Nanotechnology Research and Education Centre of the Russian Academy of Sciences, 194021 St.\ Petersburg, Russia}
\affiliation{$^3$St.\ Petersburg State University, 7/9 Universitetskaya nab., St.\ Petersburg, 199034
Russia}

\date{\today}

\begin{abstract}

We theoretically address the problem of cubic B20 helimagnets with small concentration ${c \ll 1}$ of defect bonds in external magnetic field $\bf H$, which is relevant to mixed B20 compounds at small dopant concentrations. We assume that Dzyaloshinskii-Moriya interaction and the exchange coupling constant are changed on imperfect bonds which leads to distortion of the conical spiral ordering. In one-impurity problem, we find that the distortion of the spiral pitch is long-ranged and it is governed by the Poisson equation for an electric dipole. The variation of the cone angle is described by the screened Poisson equation for two electric charges with the screening length being of the order of the spiral period. We calculate corrections to the spiral vector and to the cone angle at finite $c$. The correction to the spiral vector is shown to be independent of $H$. We demonstrate that diffuse neutron scattering caused by disorder appears in the elastic cross section as power-law decaying tails centered at magnetic Bragg peaks.

\end{abstract}


\maketitle

\section{Introduction}

More than fifty years ago it was shown that lacking of inversion center between two magnetic ions leads to an antisymmetric exchange called Dzyaloshinskii-Moriya interaction (DMI). \cite{dzyal1958,moriya1960} Competition between Heisenberg symmetrical exchange interactions and DMI can result in spiral magnetic structures. \cite{dzyal1964} Despite many years passed since the spiral magnetic ordering was observed for the first time in
magnets with DMI, helimagnets attract great interest in the present time. This interest is stimulated, in particular, by rich phase diagrams and exotic spin structures which arise due to DMI in some helimagnets. Phases with such topological states as chiral soliton lattices~\cite{togawa} (e.g., in layered helimagnet CrNb$_3$S$_6$) and skyrmion lattices consisting of close-packed magnetic vertices in B20 cubic chiral magnets (e.g., in MnSi) \cite{muhlbauer} are widely discussed now. These materials are attractive not only from a fundamental but also from a technological point of view owing to their potential applications in spintronic devices.

Such mixed B20 compounds as Mn$_{1-x}$Fe$_x$Ge and Fe$_{1-x}$Co$_x$Ge attract significant attention now due to some interesting features. For instance, a quantum critical point was observed~\cite{bauer2010} in Mn$_{1-x}$Fe$_x$Si at certain $x^*$ which separates phases with long and short range orderings~\cite{glushkov2015}. Then, it was found experimentally that the spiral vector $\mathbf{q}$ depends on the dopant concentration $x$~\cite{grig2013,grig2015}. At some critical value $x_c$, it becomes zero and $\bf q$ reverses upon $x$ variation across this point. This effect arises due to different relations between structural and magnetic chiralities of the pure compounds with $x=0$ and $x=1$~\cite{grig2013,kikuchi2016}. Microscopically, this behavior should be a result of a spin interaction change near the dopant ions. At $x\ll1$ and $x \approx 1$, dopant ions can be naturally considered as defects inside an ordered matrix, and the system can be described by a model of B20 helimagnet with small amount of defect bonds~\footnote{This model lies in agreement with the recent ESR experiments which show localized nature of Mn ions magnetic moments in MnSi~\cite{demishev2011}.}.  In our recent paper~\cite{utesov2015} we discussed in detail the spiral phase at zero magnetic field in this model. We found a remarkable electrostatic analogy in this problem: the distortion of the spiral ordering by a single defect is given by a Poisson equation for an electric dipole. The electrostatic analogy allowed us to find that a small concentration $c$ of impurities leads to a correction to the spiral vector proportional to $c$ as it was recently observed~\cite{grig2018,kinder2018} in Mn$_{1-x}$Fe$_x$Si at small $x$.

In the present paper, we extend our previous analysis~\cite{utesov2015} by introducing an external magnetic field $\bf H$. We find that the distortion of the conical spiral ordering by a single defect is described now by two coupled equations. The first equation governs the variation of the spiral pitch at each site which arose also in our previous research~\cite{utesov2015} and has the form of the Poisson equation for an electric dipole. The second equation describes the variation of the cone angle and is governed by the screened Poisson equation for two electric charges, where the ``screening length'' is of the order of the spiral period. At finite but small defects concentration, we calculate a correction to the spiral vector and find that it does not depend on $H$. A correction to the conical angle is also derived. We analyze the impact of the disorder on elastic neutron scattering. It is shown that Bragg peaks acquire power-law decaying tails due to a diffuse scattering caused by impurities.

The rest of the paper is organized as follows. In Sec.~\ref{Pure}, we discuss the model and the technique we use in our calculations. In Sec.~\ref{Disorder}, we present our analysis of B20 helimagnets with defect bonds in the external magnetic field. Sec.~\ref{Conc} contains a summary of results and our conclusions.

\section{Pure B20 helimagnet}
\label{Pure}

\begin{figure}[t]
  \centering
  \includegraphics[width=8cm]{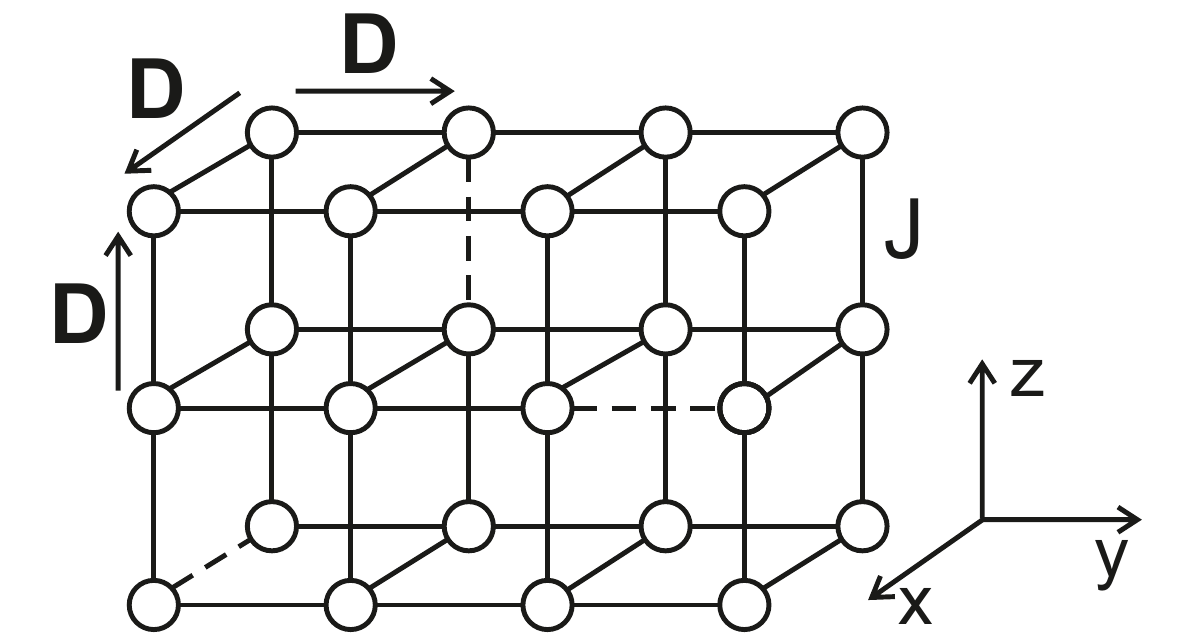}\\
  \caption{Cubic B20 helimagnet with defect bonds (dashed lines) randomly distributed over the crystal. There are defect bonds with all three spatial orientations. Exchange constant $J$ and DMI vectors $\bf D$ are also shown which values differ on solid and dashed bonds.
	\label{B20}
	}
\end{figure}

We use below the well-known Bak-Jensen model of a cubic B20 helimagnet~\cite{Bak} in the form proposed in Ref.~\cite{yi2009}. The system without defects is described by the following Hamiltonian which includes the exchange interaction, DMI, the anisotropy and the Zeeman energy:
\begin{eqnarray}
\label{ham0}
  \mathcal{H}_0 &=& \mathcal{H}_{ex} + \mathcal{H}_{dm} + \mathcal{H}_{an} + \mathcal{H}_z, \\
  \mathcal{H}_{ex} &=& -\frac12 \sum_{\mathbf{R},\mathbf{R}^\prime} J_{\mathbf{R}\mathbf{R}^\prime} (\mathbf{S}_\mathbf{R} \cdot \mathbf{S}_{\mathbf{R}^\prime}), \nonumber\\
  \mathcal{H}_{dm} &=& -\frac12 \sum_{\mathbf{R},\mathbf{R}^\prime} \mathbf{D}_{\mathbf{R}\mathbf{R}^\prime} \cdot \left[ \mathbf{S}_\mathbf{R} \times \mathbf{S}_{\mathbf{R}^\prime} \right], \nonumber\\
  \mathcal{H}_{an} &=& K  \sum_{\mathbf{R}} \left( \left(S^x_\mathbf{R}\right)^4 + \left(S^y_\mathbf{R}\right)^4 + \left(S^z_\mathbf{R}\right)^4 \right) \nonumber \\
  \mathcal{H}_z &=& \sum_{\mathbf{R}} \mathbf{h} \cdot \mathbf{S}_\mathbf{R} \nonumber,
\end{eqnarray}
where $\mathbf{h}=g \mu_B \mathbf{H}$. Without loss of generality, we consider only the nearest neighbor ferromagnetic exchange interaction with the constant $J>0$. DMI arises between nearest spins and $\mathbf{D}_{\mathbf{R}\mathbf{R}^\prime}$ is parallel to $\mathbf{R}-\mathbf{R}^\prime$ (see Fig.~\ref{B20}). We assume below the standard situation when the following hierarchy of the model constants holds: $J \gg |D| \gg |K|.$

Notice that the anisotropy can be introduced to the Hamiltonian either in the form of an exchange anisotropy or in the single-site form (as in Eq.~\eqref{ham0}). It is easy to show, however, that the anisotropy energy depends and does not depend on the spiral vector magnitude in the case of the exchange and the single-site anisotropies, respectively. As it is shown below, defects change the spiral vector. On the other hand, the critical field $H_{c1}$ of a transition from the helical state to the conical one, which is determined by anisotropic interactions, was shown experimentally (see Ref.~\cite{grig2015}) to be almost independent of the dopant concentration. That is why it is reasonable to consider the single-site cubic anisotropy as the main anisotropic interaction in the system. It is the anisotropy that can describe also destroying of the spiral ordering in pure and mixed B20 helimagnets. \cite{nakanishi1980, grig2015}

Following Refs.~\cite{kaplan,maleyev}, we introduce a local orthogonal basis at each site
\begin{eqnarray}
  \label{bas1}
  \hat{\zeta}_\mathbf{R} &=& (\hat{\mathfrak a} \cos{\mathbf{q}\mathbf{R}} + \hat{\mathfrak b} \sin{\mathbf{q}\mathbf{R}}) \cos{\alpha} + \hat{\mathfrak c} \sin{\alpha},\nonumber \\
	\hat{\eta}_\mathbf{R} &=&  - \hat{\mathfrak a} \sin{\mathbf{q}\mathbf{R}} + \hat{\mathfrak b} \cos{\mathbf{q}\mathbf{R}}, \\
 \hat{\xi}_\mathbf{R} &=& - (\hat{\mathfrak a} \cos{\mathbf{q}\mathbf{R}} + \hat{ \mathfrak b} \sin{\mathbf{q}\mathbf{R}}) \sin{\alpha} + \hat{\mathfrak c} \cos{\alpha},\nonumber
\end{eqnarray}
where $\hat{\mathfrak a}$, $\hat{\mathfrak b}$, and $\hat{\mathfrak c}$ are unit, mutually orthogonal, vectors and $\alpha$ is the cone angle ($\alpha=0$ for a plane spiral). Spin at the site $\mathbf{R}$ is expressed as
\begin{equation}
\label{spin1}
  \mathbf{S}_\mathbf{R} = S_\mathbf{R}^\zeta \hat{\zeta}_\mathbf{R} + S_\mathbf{R}^\eta \hat{\eta}_\mathbf{R} + S_\mathbf{R}^\xi \hat{\xi}_\mathbf{R}
\end{equation}
and we use the Holstein-Primakoff transformation~\cite{holstein1940} for the spin components
\begin{eqnarray}
\label{spinrep1}
  S^\zeta_\mathbf{R} &=& S - a^+_\mathbf{R} a_\mathbf{R}, \nonumber \\
  S^\eta_\mathbf{R} &\simeq& \sqrt{\frac{S}{2}} \left( a_\mathbf{R} + a^+_\mathbf{R} \right), \\
  S^\xi_\mathbf{R} &\simeq& i \sqrt{\frac{S}{2}} \left( a^+_\mathbf{R} - a_\mathbf{R} \right),\nonumber
\end{eqnarray}
where the square roots a replaced by unity.

Previous analysis \cite{grig2015} of the classical energy of the model \eqref{ham0} shows that $\mathbf{q} \parallel \hat{\mathfrak c}$ and $q=D/J$ in the helical phase. Besides, $\hat{\mathfrak c} \parallel (111)$ for $K>0$ and $\hat{ \mathfrak c} \parallel (100)$ for $K<0$ at $H=0$. A competition arises between the anisotropy and an arbitrary directed magnetic field because $\hat{\mathfrak c} \parallel \mathbf{H}$ at $K=0$ (the cone axis defined by $\hat{\mathfrak c}$ is parallel to $\bf H$). We consider the conical spiral phase with $\hat{\mathfrak c} \parallel \mathbf{H}$ thus assuming that the Zeeman energy overcomes the anisotropy which we neglect below.

By substituting Eqs.~\eqref{bas1}--\eqref{spinrep1} to Eq.~\eqref{ham0} we obtain for the term in the Hamiltonian not containing Bose-operators (i.e., for the classical energy $E$)
\begin{multline}
\label{clas1}
  \frac{E}{N} = - J S^2 \left( 3 -  \frac{q^2}{2} \cos^2{\alpha}  \right) - DS^2  (\mathbf{q} \cdot \hat{\mathfrak c}) \cos^2{\alpha} \\
	- h S \sin{\alpha},
\end{multline}
where we put the lattice parameter to be equal to unity.
Henceforth we retain only terms up to the second order in small parameter $D/J$. Minimization of Eq.~\eqref{clas1} with respect to $\bf q$ and $\alpha$ yields
\begin{eqnarray}
\label{qsp0}
  {\bf q} &=& \hat{\mathfrak c}\frac{D}{J}, \\
\label{Cone0}
  \sin{\alpha} &=& \frac{h}{h_{c2}^{(0)}}, \quad h_{c2}^{(0)} = S D^2/J,
\end{eqnarray}
where $h_{c2}^{(0)}$ is the critical field of the second order transition to the fully saturated phase.

By virtue of conditions \eqref{qsp0} and \eqref{Cone0}, terms linear in Bose-operators cancel each other in the Hamiltonian \eqref{ham0}. For terms bilinear in Bose-operators, we obtain within the second order in $D/J$
\begin{eqnarray}
\mathcal{H}^{(2)} &=& \mathcal{H}^{(2)}_{ex} + \mathcal{H}^{(2)}_{dm} + \mathcal{H}^{(2)}_z,\\
\label{hex}
  \mathcal{H}^{(2)}_{ex} &=& S J \sum_{\nu, \mathbf{R}} \Biggl[ 2 a^+_\mathbf{R} a_\mathbf{R} \left( 1 - \frac{q^2_\nu}{2} \cos^2{\alpha} \right)  \\ &-& (a^+_\mathbf{R} a_{\mathbf{R}+\mathbf{e}_\nu}+a^+_{\mathbf{R}+\mathbf{e}_\nu} a_\mathbf{R}) \left( 1 - \frac{q^2_\nu}{4} (1+\sin^2{\alpha}) \right) \nonumber \\
   &+&  (a^+_\mathbf{R} a^+_{\mathbf{R}+\mathbf{e}_\nu}+a_\mathbf{R} a_{\mathbf{R}+\mathbf{e}_\nu})\frac{q^2_\nu}{4} \cos^2{\alpha} \Biggr] \nonumber,
\end{eqnarray}
where $\nu=x,y,z$ denote components in the coordinate system shown in Fig.~\ref{B20}, $\mathbf{e}_\nu$ are vectors of elementary translations,
\begin{eqnarray}
\label{hdm}
  \mathcal{H}^{(2)}_{dm} &=&  S D \sum_{\nu, \mathbf{R}} q_\nu (\mathbf{e}_\nu \cdot \hat{\mathfrak c}) \Biggl[ 2 a^+_\mathbf{R} a_\mathbf{R} \cos^2{\alpha} \nonumber    \\
  &-& (a^+_\mathbf{R} a_{\mathbf{R}+\mathbf{e}_\nu}+a^+_{\mathbf{R}+\mathbf{e}_\nu} a_\mathbf{R}) \frac{1+\sin^2{\alpha}}{2} \nonumber \\
  &-& (a^+_\mathbf{R} a^+_{\mathbf{R}+\mathbf{e}_\nu}+a_\mathbf{R} a_{\mathbf{R}+\mathbf{e}_\nu}) \frac{\cos^2{\alpha}}{2} \Biggr],\\
\label{hz}
  \mathcal{H}^{(2)}_z &=& h \sin\alpha \sum_{\mathbf{R}} a^+_\mathbf{R} a_\mathbf{R}.
\end{eqnarray}
We omit in Eq.~\eqref{hdm} the so-called umklapp terms (see, e.g., Ref.~\cite{maleyev}), because our calculations show that their impact on the ground state properties is negligible.

One obtains at $h<h_{c2}^{(0)}$ from Eqs.~\eqref{qsp0}--\eqref{hz}
\begin{eqnarray}
\label{Ham1}
    \mathcal{H}^{(2)} &=& J S \sum_{\nu, \mathbf{R}} \Biggl[ 2 a^+_\mathbf{R} a_\mathbf{R} \left( 1 + \frac{{\mathfrak c}^2_\nu q^2}{2} \cos^2{\alpha} \right)  \\ &-& (a^+_\mathbf{R} a_{\mathbf{R}+\mathbf{e}_\nu}+a^+_{\mathbf{R}+\mathbf{e}_\nu} a_\mathbf{R}) \left( 1 + \frac{{\mathfrak c}^2_\nu q^2}{4} (1+\sin^2{\alpha}) \right) \nonumber \\
   &+&  (a^+_\mathbf{R} a^+_{\mathbf{R}+\mathbf{e}_\nu}+a_\mathbf{R} a_{\mathbf{R}+\mathbf{e}_\nu})\frac{{\mathfrak c}^2_\nu q^2}{4} \cos^2{\alpha} \Biggr] \nonumber.
\end{eqnarray}
After the Fourier transform, bilinear Hamiltonian \eqref{Ham1} yields the well-known classical spectrum
\begin{equation}
\label{Spec1}
  \varepsilon(\mathbf{k}) = S |D| k \cos{\alpha}.
\end{equation}
In the fully saturated phase (i.e., at $h \geq h^{(0)}_{c2}$), $\cos{\alpha}=0$ and the spin-wave spectrum
\begin{equation}
\label{Spec2}
  \varepsilon(\mathbf{k}) = h - h_{c2} + S J k^2
\end{equation}
acquires a gap.

\section{Disordered system}
\label{Disorder}

We introduce now a disorder in model \eqref{ham0} by changing $J_{\mathbf{R}\mathbf{R}^\prime}$ and ${\bf D}_{\mathbf{R}\mathbf{R}^\prime}$ (so that it remains parallel to the bond) on some randomly distributed bonds which concentration in the system is $c$. The model constants are $J^\prime=J + u_{ex}$ and $D^\prime = D + u_{dm}$ on defect bonds. We assume below only the smallness of $c\ll1$ whereas $|u_{dm}|$ and $|u_{ex}|$ can be of the order of $D$ and $J$, respectively.

Let us begin with the one-impurity problem and consider a single defect bond between sites $\mathbf{R}_0 = (0,0,0)$ and $\mathbf{R}_1 = (0,0,1)$. Corresponding perturbation to the Hamiltonian reads as
\begin{multline}
\label{Pert1}
  V = V_{ex} + V_{dm} \\
	=- u_{ex} \mathbf{S}_{\mathbf{R}_0} \cdot \mathbf{S}_{\mathbf{R}_1} - u_{dm} \mathbf{e}_z \cdot \left[ \mathbf{S}_{\mathbf{R}_0} \times \mathbf{S}_{\mathbf{R}_1} \right] .
\end{multline}
The analysis below shows that one needs to take into account only terms up to the first order in $D/J$ in Eq.~\eqref{Pert1} to obtain accurate results in the leading order in $D/J$. Then, we have for the perturbation from Eqs.~\eqref{bas1}--\eqref{spinrep1}, \eqref{qsp0}, and \eqref{Pert1}
\begin{eqnarray}
\label{vdm}
  V_{dm} &=& - S u_{dm} {\mathfrak c}_z \sqrt{\frac{S}{2}}  \Bigl[ ( a^+_{\mathbf{R}_1} + a_{\mathbf{R}_1} - a^+_{\mathbf{R}_0} - a_{\mathbf{R}_0} ) \cos{\alpha} \nonumber \\
  &-&  i {\mathfrak c}_z q ( a^+_{\mathbf{R}_1} - a_{\mathbf{R}_1} + a^+_{\mathbf{R}_0} - a_{\mathbf{R}_0} ) \cos{\alpha} \sin{\alpha} \Bigr] \\
  &+& i S u_{dm} {\mathfrak c}_z (a^+_{\mathbf{R}_0} a_{\mathbf{R}_1} - a^+_{\mathbf{R}_1} a_{\mathbf{R}_0} ) \sin{\alpha}, \nonumber\\
\label{vex}
  V_{ex} &=& S u_{ex} {\mathfrak c}_z q \sqrt{\frac{S}{2}} \Bigl[ ( a^+_{\mathbf{R}_1} + a_{\mathbf{R}_1} - a^+_{\mathbf{R}_0} - a_{\mathbf{R}_0} ) \cos{\alpha} \nonumber \\
  &-&  i \frac{{\mathfrak c}_z q}{2} ( a^+_{\mathbf{R}_1} - a_{\mathbf{R}_1} + a^+_{\mathbf{R}_0} - a_{\mathbf{R}_0} ) \cos{\alpha} \sin{\alpha} \Bigr] \\
  &+& S u_{ex} \Bigl[ ( a^+_{\mathbf{R}_0} a_{\mathbf{R}_0} + a^+_{\mathbf{R}_1} a_{\mathbf{R}_1} - a^+_{\mathbf{R}_1} a_{\mathbf{R}_0} - a^+_{\mathbf{R}_0} a_{\mathbf{R}_1}) \nonumber \\
  &+& i {\mathfrak c}_z q (a^+_{\mathbf{R}_1} a_{\mathbf{R}_0} - a^+_{\mathbf{R}_0} a_{\mathbf{R}_1} )  \sin{\alpha} \Bigr]. \nonumber
\end{eqnarray}

\begin{figure}
  \centering
  \includegraphics[height=2.cm]{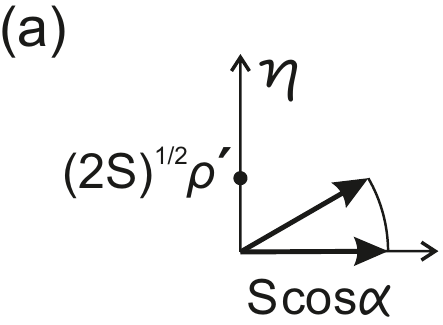}
  \includegraphics[height=2.cm]{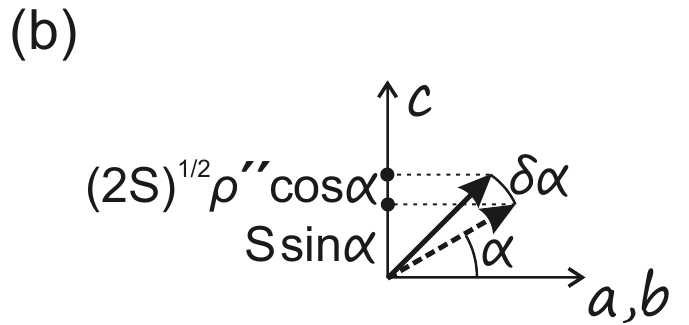}\\
  \caption{Parameters $\rho'_\mathbf{R}$ and $\rho''_\mathbf{R}$ arising in Eqs.~\eqref{Shift1} due to the defects lead, correspondingly, to (a) additional spin rotation in the spiral plane shown by arrows and (b) to a correction $\delta \alpha$ to the cone angle $\alpha$.
}\label{Rotation}
\end{figure}

One can see from Eqs.~\eqref{vdm} and \eqref{vex} that defect bonds provide linear terms in Bose-operators. This signifies that the bare ground state (conical spiral) is modified by impurities. To eliminate the linear terms in the  Hamiltonian, we perform the following shift in Bose-operators:
\begin{eqnarray}
\label{Shift1}
  a_\mathbf{R} &=& b_\mathbf{R} + \rho_\mathbf{R} = b_\mathbf{R} + \rho^\prime_\mathbf{R} + i \rho^{\prime\prime}_\mathbf{R}, \\
  a^+_\mathbf{R} &=& b^+_\mathbf{R} + \rho^*_\mathbf{R} = b^+_\mathbf{R} + \rho^\prime_\mathbf{R} - i \rho^{\prime\prime}_\mathbf{R}, \nonumber
\end{eqnarray}
where real parameters $\rho^\prime_\mathbf{R}$ and $\rho^{\prime\prime}_\mathbf{R}$ has to be chosen such that linear terms in Bose-operators vanish in the Hamiltonian.  Notice that spin representation~\eqref{spinrep1} with the truncated square roots can be used only if the ``condensate densities'' are small, i.e., if $|\rho_\mathbf{R}| \ll \sqrt{S}$. The physical meaning of $\rho^\prime_\mathbf{R}$ and $\rho^{\prime\prime}_\mathbf{R}$ can be revealed from an analysis of Eqs.~\eqref{bas1}--\eqref{spinrep1} and \eqref{Shift1}. It is easy to show that an additional rotation of spins arises at site $\bf R$ in the spiral plane (i.e., in the plane in which spins rotate) by an angle having the form (see also Fig.~\ref{Rotation}(a))
\begin{equation}
\label{rp}
	\rho^\prime_\mathbf{R} \frac{1}{\cos{\alpha}}\sqrt{\frac2S}.
\end{equation}
Appearance of $\rho^{\prime\prime}_\mathbf{R}$ leads to a correction $\delta\alpha_{\bf R}$ to the conical angle $\alpha$ which can be found from the equation
\begin{equation}
\label{Ang1}
  S \sin(\alpha + \delta \alpha_{\bf R}) = (S-(\rho_{\bf R}^{\prime})^2 - (\rho_{\bf R}^{\prime\prime})^2)\sin \alpha + \sqrt{2S} \rho_{\bf R}^{\prime\prime}  \cos \alpha.
\end{equation}
It is seen from the results below that the last term in Eq.~\eqref{Ang1} is much greater than $(\rho_{\bf R}^{\prime})^2 + (\rho_{\bf R}^{\prime\prime})^2 $ by the parameter $D/J$ at not very large field (i.e., at $\cos\alpha\sim1$). As a result, one obtains from Eq.~\eqref{Ang1}
\begin{equation}
\label{rpp}
\delta \alpha_{\bf R} = \rho_{\bf R}^{\prime\prime}\sqrt{\frac2S}
\end{equation}
(see also Fig.~\ref{Rotation}(b) for an illustration).

\subsection{Defects in DMI only}

We begin with the technically simpler case of $u_{ex}=0$ and $u_{dm}\ne0$. One infers from Eqs.~\eqref{Ham1}, \eqref{vdm}--\eqref{Shift1} that the following system of equations should hold on every site in order to eliminate terms linear in Bose-operators:
\begin{eqnarray}
\label{Sys1}
  && J \sum_{\nu} \Biggl[ (2 \rho^\prime_\mathbf{R} - \rho^\prime_{\mathbf{R} + \mathbf{e}_\nu} - \rho^\prime_{\mathbf{R} - \mathbf{e}_\nu})\left( 1 + \frac{{\mathfrak c}^2_\nu q^2}{2}\right) \\ && + 2 i \rho^{\prime\prime}_\mathbf{R} \left( 1 + \frac{{\mathfrak c}^2_\nu q^2}{2}\right) \nonumber\\
  &&  - i(\rho^{\prime\prime}_{\mathbf{R} + \mathbf{e}_\nu} + \rho^{\prime\prime}_{\mathbf{R} - \mathbf{e}_\nu}) \left( 1 + \frac{{\mathfrak c}^2_\nu q^2 \sin^2{\alpha}}{2}\right) \Biggr]  \nonumber \\
  && = u_{dm} {\mathfrak c}_z  \left[ \sqrt{\frac{S}{2}} \cos{\alpha} (1 - i {\mathfrak c}_z q \sin{\alpha})  + i\rho_{\mathbf{R}_0} \sin{\alpha} \right] \delta_{\mathbf{R}, \mathbf{R}_1}, \nonumber \\
   && + u_{dm} {\mathfrak c}_z  \left[ - \sqrt{\frac{S}{2}}  \cos{\alpha} (1 + i {\mathfrak c}_z q \sin{\alpha})  - i \rho_{\mathbf{R}_1} \sin{\alpha} \right] \delta_{\mathbf{R}, \mathbf{R}_0}. \nonumber
\end{eqnarray}
In the absence of the ``source'' in the right hand side (i.e., at $u_{dm}=0$), Eqs.~\eqref{Sys1} yield $\rho_\mathbf{R} \equiv 0$. One concludes that the ``source'' provides $\rho^\prime \sim u_{dm}/J$ and $\rho^{\prime\prime} \sim (u_{dm}/J)(D/J)$. The real part of Eqs.~\eqref{Sys1} reads in the leading order in $D/J$ as
\begin{multline}
\label{Re1}
\sum_{\nu} (2 \rho^\prime_\mathbf{R} - \rho^\prime_{\mathbf{R} + \mathbf{e}_\nu} - \rho^\prime_{\mathbf{R} - \mathbf{e}_\nu}) \\
	= (\delta_{\mathbf{R}, \mathbf{R}_1}-\delta_{\mathbf{R}, \mathbf{R}_0}) \frac{u_{dm}}{J} {\mathfrak c}_z \sqrt{\frac{S}{2}} \cos{\alpha}.
\end{multline}
One recognizes in Eq.~\eqref{Re1} a discrete (lattice) variant of a Poisson equation for an electric dipole which reads in the continuous limit as
\begin{equation}
\label{Re2}
  \triangle \rho^\prime = - \frac{u_{dm} {\mathfrak c}_z \cos{\alpha} }{J} \sqrt{\frac{S}{2}} \left[ \delta(\mathbf{r} - \mathbf{R}_1) - \delta(\mathbf{r} - \mathbf{R}_0) \right].
\end{equation}
The well-known solution of Eq.~\eqref{Re2} has the form
\begin{equation}
\label{Re3}
  \rho^\prime(\mathbf{r}) =  \frac{u_{dm} {\mathfrak c}_z \cos{\alpha} }{ 4 \pi J} \sqrt{\frac{S}{2}} \left( \frac{1}{|\mathbf{r} - \mathbf{R}_1|} - \frac{1}{|\mathbf{r} - \mathbf{R}_0|} \right).
\end{equation}

If the concentration of defects is finite, we have a system with randomly distributed electric dipoles. Using the electrostatic superposition principle, one concludes that the contribution to average ``electric polarization'' per unit volume $\mathbf{P}$ from each of these dipole is proportional to ${\mathfrak c}_\nu$ so that we have
\begin{equation}
\label{Pol1}
  \mathbf{P} = \hat{{\mathfrak c}} \frac{c}{3} \frac{ u_{dm} \cos{\alpha} }{4 \pi J}\sqrt{\frac{S}{2}}.
\end{equation}
Then, we find using the well-known electrostatic relation $\overrightarrow{\nabla} \rho^\prime = 4 \pi \mathbf{P}$
\begin{equation}
\label{ReSol}
  \langle \rho^\prime(\mathbf{r}) \rangle  = (\mathbf{r} \cdot \hat{{\mathfrak c}})\frac{c}{3} \frac{ u_{dm} \cos{\alpha} }{J} \sqrt{\frac{S}{2}}
\end{equation}
that results in a correction to the spiral vector $\delta \mathbf{q} \parallel \mathbf{q}$. Using Eqs.~\eqref{rp} and~\eqref{ReSol}, one obtains
\begin{equation}
\label{Spir1}
  \delta q = \frac{c}{3} \frac{ u_{dm} }{ J}
\end{equation}
that is independent of the magnetic field. Such a linear in dopant concentration correction to the spiral vector was observed recently experimentally in Mn$_{1-x}$Fe$_x$Si. \cite{grig2018,kinder2018}

It should be pointed out that solution~\eqref{ReSol} lies in contradiction with the requirement $|\rho^\prime_\mathbf{R}| \ll \sqrt{S}$ at large enough $R$. It happens because the finite concentration of defects changes the spiral vector. As a result, orientation of some spins differs significantly from their orientation in the pure system. However, Eq.~\eqref{Spir1} for the correction to the spiral vector is correct. To show this, one has to repeat the above derivations for finite $c$ trying the spiral vector in the form $q' = q + \delta q$ from the very beginning. This leads to additional ``charges'' in system~\eqref{Sys1} whose values are proportional to $\delta q$. As a result, one obtains $\bf P=0$ and $\langle \rho^\prime_\mathbf{R} \rangle =0$ instead of Eqs.~\eqref{Pol1} and \eqref{ReSol} if the value of $\delta q$ is given by Eq.~\eqref{Spir1} (see also Ref.~\cite{utesov2015} for more details). Then, we conclude again (not violating the requirement $|\rho^\prime_\mathbf{R}| \ll \sqrt{S}$) that in average defects result in correction~\eqref{Spir1} to the spiral pitch~\eqref{qsp0}.

Now we turn to the imaginary part of Eqs.~\eqref{Sys1} which can be rewritten in the form
\begin{multline}
\label{Im1}
q^2 \cos^2{\alpha} \rho^{\prime\prime}_\mathbf{R}  \\
+  \sum_{\nu} \left( 1 + \frac{{\mathfrak c}^2_\nu q^2 \sin^2{\alpha}}{2}\right)
[ 2 \rho^{\prime\prime}_\mathbf{R} - \rho^{\prime\prime}_{\mathbf{R} + \mathbf{e}_\nu} - \rho^{\prime\prime}_{\mathbf{R} - \mathbf{e}_\nu} ] \\
= - \frac{u_{dm}}{J} {\mathfrak c}_z \sin{\alpha}
\Biggl[ \sqrt{\frac{S}{2}}  {\mathfrak c}_z q \cos{\alpha}
 (\delta_{\mathbf{R}, \mathbf{R}_1}+\delta_{\mathbf{R}, \mathbf{R}_0})  \\
- \rho^\prime_{\mathbf{R}_0} \delta_{\mathbf{R}, \mathbf{R}_1} + \rho^\prime_{\mathbf{R}_1} \delta_{\mathbf{R}, \mathbf{R}_0}\Biggr] .
\end{multline}
It follows from Eqs.~\eqref{Im1} that $\rho^{\prime\prime}_\mathbf{R}\equiv0$ at $h=0$ (i.e., at $\sin\alpha=0$). It means that the spiral remains plane at $h=0$ and Eq.~\eqref{Spir1} recovers the result of our previous consideration~\cite{utesov2015} devoted to B20 helimagnets at zero field.

At $h\ne0$, $\rho^\prime_{\mathbf{R}_0}$ and $\rho^\prime_{\mathbf{R}_1}$ arising in Eqs.~\eqref{Im1} can be found from Eqs.~\eqref{Re1} for $\mathbf{R}=\mathbf{R}_0$ and $\mathbf{R}=\mathbf{R}_1$ which have the form
\begin{eqnarray}
\label{sys2}
\left\{
  \begin{array}{ll}
    6 \rho^\prime_{\mathbf{R}_0} - 4 \rho^\prime_{\mathbf{R}_0 + \mathbf{e}_x} - \rho^\prime_{\mathbf{R}_1} - \rho^\prime_{\mathbf{R}_0 - \mathbf{e}_z}&= -  \frac{u_{dm}}{J} {\mathfrak c}_z \sqrt{\frac{S}{2}} \cos{\alpha}, \\
    6 \rho^\prime_{\mathbf{R}_1} - 4 \rho^\prime_{\mathbf{R}_1 + \mathbf{e}_x} - \rho^\prime_{\mathbf{R}_0} - \rho^\prime_{\mathbf{R}_1 + \mathbf{e}_z}&=   \frac{u_{dm}}{J} {\mathfrak c}_z \sqrt{\frac{S}{2}} \cos{\alpha},
  \end{array}
\right.
\end{eqnarray}
where we assume that $\rho^\prime_{\mathbf{R}_0 \pm \mathbf{e}_x}=\rho^\prime_{\mathbf{R}_0 \pm \mathbf{e}_y}$ and $\rho^\prime_{\mathbf{R}_1 \pm \mathbf{e}_x}=\rho^\prime_{\mathbf{R}_1 \pm \mathbf{e}_y}$ by symmetry. By numerical solution of Eqs.~\eqref{Re1} we observe that Eq.~\eqref{Re3} obtained in the continuum limit starts working well right from sites neighboring to the defect bond in a broad range of parameters $|u_{dm}|\alt D$ (see also Ref.~\cite{utesov2015}). Then, using Eq.~\eqref{Re3} for $\rho^\prime_{\mathbf{R}_{0,1} + \mathbf{e}_x}$, $\rho^\prime_{\mathbf{R}_{0} - \mathbf{e}_z}$ and $\rho^\prime_{\mathbf{R}_1 + \mathbf{e}_z}$ in Eqs.~\eqref{sys2} and subtracting the first equation \eqref{sys2} from the second one, we find
\begin{equation}
\label{Re4}
  \rho^\prime_{\mathbf{R}_1} - \rho^\prime_{\mathbf{R}_0} = \frac{2 u_{dm} {\mathfrak c}_z \sqrt{\frac{S}{2}} \cos{\alpha}}{7 J}  \frac{4(1-1/\sqrt{2})+1/2 + 4 \pi}{4\pi}.
\end{equation}
Also taking into account that $\rho^\prime_{\mathbf{R}_1}=-\rho^\prime_{\mathbf{R}_0}$ one obtains from Eq.~\eqref{Re4}
\begin{equation}
\label{Re5}
\rho^\prime_{\mathbf{R}_1} = -\rho^\prime_{\mathbf{R}_0} \approx 0.16 \frac{u_{dm}}{J} {\mathfrak c}_z \sqrt{\frac{S}{2}} \cos{\alpha}.
\end{equation}
Using Eq.~\eqref{Re5}, we come from Eqs.~\eqref{Im1} to the following equation in the continuous limit:
\begin{multline}
\label{Im2}
  \triangle \rho^{\prime\prime} - q^2 \cos^2{\alpha} \rho^{\prime\prime} = \frac{u_{dm}}{J} {\mathfrak c}^2_z \sqrt{\frac{S}{2}} \cos{\alpha} \sin{\alpha} \\
  \times \left( q + 0.16 \frac{u_{dm}}{J} \right) [\delta(\mathbf{r} - \mathbf{R}_1)+\delta(\mathbf{r} - \mathbf{R}_0)],
\end{multline}
which is a screened Poisson equation for two point charges (cf.\ Eq.~\eqref{Re2}). The well-known solution of this equation has the form
\begin{multline}
\label{Im3}
  \rho^{\prime\prime}(\mathbf{r}) = -  \frac{u_{dm}}{4 \pi J} {\mathfrak c}^2_z \sqrt{\frac{S}{2}} \cos{\alpha} \sin{\alpha} \left( q + 0.16 \frac{u_{dm}}{J} \right) \\
	\times \left[ \frac{e^{- |q|  |\mathbf{r} - \mathbf{R}_1|\cos{\alpha}}}{|\mathbf{r} - \mathbf{R}_1|} + \frac{e^{- |q|  |\mathbf{r} - \mathbf{R}_0| \cos{\alpha}}}{|\mathbf{r} - \mathbf{R}_0|}\right].
\end{multline}
One can see that the quantity
\begin{equation}\label{Leng1}
  \lambda = \frac{1}{|q| \cos{\alpha}}
\end{equation}
plays a role of the ``screening length''.

For finite defect concentration $c$ we use the superposition principle of electrostatics and perform averaging over disorder configurations utilizing formula $\int d^3 r e^{- r/\lambda}/r = 4 \pi \lambda^2$. As a result we obtain
\begin{equation}
\label{Im4}
  \langle \rho^{\prime\prime} \rangle = - \frac{2 c u_{dm} }{3J q^2} \sqrt{\frac{S}{2}} \left( q + 0.16 \frac{u_{dm}}{J} \right) \frac{\sin{\alpha}}{\cos{\alpha}},
\end{equation}
that leads to a correction $\delta\alpha$ to the cone angle $\alpha$. We derive from Eqs.~\eqref{rpp} and \eqref{Im4}
\begin{equation}
\label{Ang2}
  \delta \alpha = - \frac{2 c u_{dm} }{3J q^2} \left( q + 0.16 \frac{u_{dm}}{J} \right) \frac{\sin{\alpha}}{\cos{\alpha}}.
\end{equation}

It should be noted that the correction $\delta q$ to the spiral vector could induce additional ``imaginary charges'' which would modify the solution of Eqs.~\eqref{Sys1} for $\rho^{\prime\prime}$. Thus, the following additional term arises in the left hand side of Eqs.~\eqref{Sys1}:
\begin{equation}
\label{Char1}
  - i J S \sqrt{\frac{S}{2}} \sin{\alpha}\cos{\alpha} \, \delta q^2.
\end{equation}
However, this term can be safely neglected because, according to Eq.~\eqref{Spir1}, it is proportional to $c^2$.

Notice that if the magnetic field is close to its critical value $h_{c2}$ (i.e., if $\alpha \approx \pi/2$), the screening length~\eqref{Leng1} and Eq.~\eqref{Ang2} could be infinitely large. It signifies that our consideration is invalid at high enough field.
It is well known that according to the general theorem~\cite{theorem} there should be an intermediate state called Bose-glass between the ordered and the fully saturated phases in systems with continuous symmetry. The Bose-glass is a gapless and compressible state (i.e., it has a finite susceptibility). Although DMI breaks the continuous symmetry in our system, we anticipate an emergence of a glassy phase between the ordered and the fully saturated states similar to the conventional Bose-glass. Let us consider, for example, defect bonds with $D^\prime>D$. In this case, regions with high concentration of defects become magnetically ordered earlier than the whole system upon magnetic field decreasing (cf.\ Eq.~\eqref{Spec2}). Then, the disappearance of the magnetic order upon field increasing is {\it qualitatively} similar in our system to a percolation transition as it is the case for transitions to Bose-glass phases. \cite{syromyat}

Interestingly, in Eq.~\eqref{Ang2} there is a possibility to change the sign of $\delta \alpha$ twice by varying $u_{dm}$. Let's consider $q>0$, and let's assume that $u_{dm}>0$. The in-plane angle between spins on defect bonds increases in this case so that it is harder to magnetize such bonds. Then, it is intuitively clear that such defects lead to a negative correction to the conical angle in agreement with Eq.~\eqref{Ang2}. There are two possibilities for negative $u_{dm}$. First, if $0>0.16 u_{dm}/J>- q$, the absolute value of the in-plane angle between spins on the defect bond is smaller than that in the pure system. Thus, it is easier to magnetize such bonds and the conical angle increases in accordance with Eq.~\eqref{Ang2}. Second, if $0.16 u_{dm}/J< - q$, the spins on the defect rotate in the opposite direction in comparison with the pure system and the absolute value of the angle of rotation is larger than that in the pure system (cf.\ Eq.~\eqref{Re5}). This should lead to $\rho^{\prime\prime}<0$ and, in turn, to a negative $\delta \alpha$ in agreement with Eq.~\eqref{Ang2}.

\subsection{Defects in both DMI and exchange interaction}

Let us consider now a general situation when both $u_{ex}$ and $u_{dm}$ can be nonzero. In comparison with the analysis made in the previous section, calculations become more cumbersome. The counterpart of Eqs.~\eqref{Re1} has the form
\begin{multline}
\label{Real1}
  \sum_{\nu} (2 \rho^\prime_\mathbf{R} - \rho^\prime_{\mathbf{R} + \mathbf{e}_\nu} - \rho^\prime_{\mathbf{R} - \mathbf{e}_\nu}) = (\delta_{\mathbf{R}, \mathbf{R}_1}-\delta_{\mathbf{R}, \mathbf{R}_0}) \\
\times\left[ \frac{u_{dm} - u_{ex}q}{J} {\mathfrak c}_z \sqrt{\frac{S}{2}} \cos{\alpha}- \frac{u_{ex}}{J} (\rho^\prime_{\mathbf{R}_1} - \rho^\prime_{\mathbf{R}_0}) \right].
\end{multline}
This equations can be solved self-consistently as in Ref.~\cite{utesov2015}. Denoting $\rho^\prime_{\mathbf{R}_1} - \rho^\prime_{\mathbf{R}_0} = \gamma$, the solution of Poisson equation \eqref{Real1} reads as
\begin{eqnarray}
\label{Real2}
  \rho^\prime(\mathbf{r}) &=&  \frac{(u_{dm}-u_{ex}q) {\mathfrak c}_z \cos{\alpha} \sqrt{\frac{S}{2}} - \gamma u_{ex} }{ 4 \pi J} \\ && \times \left( \frac{1}{|\mathbf{r} - \mathbf{R}_1|} - \frac{1}{|\mathbf{r} - \mathbf{R}_0|} \right). \nonumber
\end{eqnarray}
To find $\gamma$, we write down two equations from system~\eqref{Real1} for sites ${\mathbf{R}_1}$ and ${\mathbf{R}_0}$ (cf.\ Eq.~\eqref{sys2})
\begin{eqnarray}
\label{arr}
\left\{
  \begin{array}{llll}
    6 \rho^\prime_{\mathbf{R}_0} - 4 \rho^\prime_{\mathbf{R}_0 + \mathbf{e}_x} - \rho^\prime_{\mathbf{R}_1} - \rho^\prime_{\mathbf{R}_0 - \mathbf{e}_z}&= \\ -  \left[ \frac{u_{dm} - u_{ex}q}{J} {\mathfrak c}_z \sqrt{\frac{S}{2}} \cos{\alpha}  - \frac{ \gamma u_{ex}}{J}  \right], \\
    6 \rho^\prime_{\mathbf{R}_1} - 4 \rho^\prime_{\mathbf{R}_1 + \mathbf{e}_x} - \rho^\prime_{\mathbf{R}_0} - \rho^\prime_{\mathbf{R}_1 + \mathbf{e}_z}&= \\   \left[ \frac{u_{dm} - u_{ex}q}{J} {\mathfrak c}_z \sqrt{\frac{S}{2}} \cos{\alpha}  - \frac{ \gamma u_{ex}}{J}  \right].
  \end{array}
\right.
\end{eqnarray}
As in the previous section, we observe by numerical solution of Eqs.~\eqref{Real1} that Eq.~\eqref{Real2} obtained in the continuous limit starts working well right from sites neighboring to the defect bond in a broad range of parameters $|u_{dm}|\alt D$ and $|u_{ex}|\alt J$. Then, using Eq.~\eqref{Real2} for $\rho^\prime_{\mathbf{R}_{0,1} + \mathbf{e}_x}$, $\rho^\prime_{\mathbf{R}_{0} - \mathbf{e}_z}$ and $\rho^\prime_{\mathbf{R}_1 + \mathbf{e}_z}$ in Eqs.~\eqref{arr} and subtracting the first equation \eqref{arr} from the second one, we obtain
\begin{eqnarray}
\label{arr2}
  && \rho^\prime_{\mathbf{R}_1} - \rho^\prime_{\mathbf{R}_0} = \frac{2 [(u_{dm}- u_{ex}q) {\mathfrak c}_z \sqrt{\frac{S}{2}} \cos{\alpha} - \gamma u_{ex}]}{7 J} \nonumber  \\ \label{Real3} && \times \frac{4(1-1/\sqrt{2})+1/2 + 4 \pi}{4\pi} \\ &&\approx 0.16 \frac{2 [(u_{dm}- u_{ex}q) {\mathfrak c}_z \sqrt{\frac{S}{2}} \cos{\alpha} - \gamma u_{ex}]}{J} = \gamma. \nonumber
\end{eqnarray}
The last equality in Eq.~\eqref{arr2} is the condition of the self-consistency of our derivation, which gives
\begin{equation}
\label{Real4}
  \rho^\prime_{\mathbf{R}_1} - \rho^\prime_{\mathbf{R}_0} = \gamma = 0.32 \frac{ (u_{dm}- u_{ex}q) {\mathfrak c}_z \sqrt{\frac{S}{2}} \cos{\alpha}}{J\left(1+0.32u_{ex}/J\right)}.
\end{equation}
Using this expression, we obtain from Eq.~\eqref{Real2}
\begin{eqnarray}
\label{Real5}
  \rho^\prime(\mathbf{r}) &=&  \frac{{\cal Q}_r {\mathfrak c}_z \cos{\alpha} \sqrt{\frac{S}{2}}}{ 4 \pi J} \left( \frac{1}{|\mathbf{r} - \mathbf{R}_1|} - \frac{1}{|\mathbf{r} - \mathbf{R}_0|} \right), \\
 \label{Rcharge} {\cal Q}_r &=& \frac{u_{dm}- u_{ex}q}{1+0.32u_{ex}/J }.
\end{eqnarray}
As it is done in the previous section, we derive the following analog of Eq.~\eqref{Spir1}:
\begin{equation}
\label{Spiral1}
  \delta{q} = \frac{c}{3} \frac{u_{dm}- u_{ex}q}{J(1+0.32u_{ex}/J )}.
\end{equation}
One can see from Eq.~\eqref{Spiral1} that correction $\delta{q}$ is zero if $u_{dm}=u_{ex}q$. It is quite expected because this condition means that the ratio $D'/J'$ on the defect bond is equal to its value $D/J=q$ in the matrix.

The counterpart of Eqs.~\eqref{Im1} has the form
\begin{eqnarray}
\label{Imag1}
  && q^2 \cos^2{\alpha} \rho^{\prime\prime}_\mathbf{R}  +  \sum_{\nu} \left( 1 + \frac{{\mathfrak c}^2_\nu q^2 \sin^2{\alpha}}{2}\right) \\ &&\times [ 2 \rho^{\prime\prime}_\mathbf{R} - \rho^{\prime\prime}_{\mathbf{R} + \mathbf{e}_\nu} - \rho^{\prime\prime}_{\mathbf{R} - \mathbf{e}_\nu} ] = - \frac{u_{dm}- u_{ex}q /2}{J} {\mathfrak c}^2_z q \sqrt{\frac{S}{2}} \nonumber \\
&& \times  \sin{\alpha} \cos{\alpha} (\delta_{\mathbf{R}, \mathbf{R}_1} +\delta_{\mathbf{R}, \mathbf{R}_0}) \nonumber \\ && + {\mathfrak c}_z \frac{u_{dm}- u_{ex}q}{J} \sin \alpha (\rho^\prime_{\mathbf{R}_0} \delta_{\mathbf{R}, \mathbf{R}_1} - \rho^\prime_{\mathbf{R}_1} \delta_{\mathbf{R}, \mathbf{R}_0}) \nonumber \\ &&- \frac{u_{ex}}{J} (\rho^{\prime\prime}_{\mathbf{R}_1} - \rho^{\prime\prime}_{\mathbf{R}_0})(\delta_{\mathbf{R}, \mathbf{R}_1}-\delta_{\mathbf{R}, \mathbf{R}_0}), \nonumber
\end{eqnarray}
where $\rho^\prime_{\mathbf{R}_1} = - \rho^\prime_{\mathbf{R}_0} = \gamma/2$ and $\gamma$ is given by Eq.~\eqref{Real4}.
Values $\rho^{\prime\prime}_{\mathbf{R}_1}$ and $\rho^{\prime\prime}_{\mathbf{R}_0}$ appearing in the right-hand side of Eq.~\eqref{Imag1} can be found in a self-consistent manner as it is done above.
As a result, we come again to the screened Poisson equation (similar to Eq.~\eqref{Im2}) for equal charges located at $\mathbf{R}_1$ and $\mathbf{R}_0$. Its solution has the form (cf.\ Eq.~\eqref{Im3})
\begin{eqnarray}
\label{Imag2}
  \rho^{\prime\prime}(\mathbf{r}) &=& -   \frac{{\cal Q}_i {\mathfrak c}^2_z \sqrt{\frac{S}{2}} \cos{\alpha} \sin{\alpha}}{4 \pi  J} \\ &\times&  \left[ \frac{e^{- |q|  |\mathbf{r} - \mathbf{R}_1|\cos{\alpha}}}{|\mathbf{r} - \mathbf{R}_1|} + \frac{e^{- |q|  |\mathbf{r} - \mathbf{R}_0| \cos{\alpha}}}{|\mathbf{r} - \mathbf{R}_0|}\right], \nonumber \\ \label{Imag3}
  {\cal Q}_i &=&  q\left(u_{dm}-\frac{u_{ex}q}{2}\right) + \frac{0.16(u_{dm}- u_{ex}q)^2}{(1 + 0.32 u_{ex}/J)J}.
\end{eqnarray}
Using Eq.~\eqref{Imag2}, we obtain for the correction to the conical angle
\begin{equation}
\label{Angle1}
  \delta \alpha = - \frac{2 c }{3 q^2} \left[ q \frac{u_{dm}- u_{ex}q /2}{J} + \frac{0.16(u_{dm}- u_{ex}q)^2}{(1 + 0.32 u_{ex}/J)J^2} \right] \frac{\sin{\alpha}}{\cos{\alpha}}.
\end{equation}
In contrast to $\delta q$, $\delta \alpha\ne0$ at $u_{dm} = u_{ex}q$ due to the first term in the brackets in Eq.~\eqref{Angle1}. This effect originates from the nonlinear dependence of the cone angle on $D$ (cf.\ Eq.~\eqref{Cone0}). Then, even if $q$ is not affected by impurities, there is a distortion in the conical angle.

\subsection{Elastic neutron scattering}

We perform an analysis similar to that presented in Ref.~\cite{utesov2015} to find the elastic neutron scattering cross section given by the general expression~\cite{Lowesey}
\begin{equation}
\label{cs}
  \frac{d \sigma}{d \Omega} \propto
	\sum_{\mathbf{R}_1,\mathbf{R}_2} e^{i \mathbf{Q} (\mathbf{R}_1-\mathbf{R}_2)}
	\sum_{\chi,\eta} (\delta_{\chi\eta}-{\widehat{Q}}^\chi\widehat{Q}^\eta)
	\langle S^\chi_{\mathbf{R}_1}\rangle \langle S^\eta_{\mathbf{R}_2}\rangle,
\end{equation}
where $\mathbf{Q}$ is the momentum transfer, $\widehat{\mathbf{Q}}=\mathbf{Q}/Q$, $\chi,\eta=\mathfrak{a},\mathfrak{b},\mathfrak{c}$, and $\langle \dots \rangle$ denotes an average over quantum and thermal fluctuations (it should not be confused with averaging over disorder configurations). Using Eqs.~\eqref{bas1}--\eqref{spinrep1} and \eqref{Shift1}, one derives the following expressions for spin components in Eq.~\eqref{cs}:
\begin{eqnarray}
  \langle S^\mathfrak{a}_\mathbf{R} \rangle &=& \left[ S - (\rho^\prime_\mathbf{R})^2 -(\rho^{\prime\prime}_\mathbf{R})^2 \right] \cos{\alpha} \cos{\mathbf{q}^\prime \mathbf{R}} \\
  &-& \sqrt{2S} \rho^\prime_\mathbf{R} \sin{\mathbf{q}^\prime \mathbf{R}} - \sqrt{2S} \rho^{\prime\prime}_\mathbf{R} \sin{\alpha} \cos{\mathbf{q}^\prime \mathbf{R}},   \nonumber \\
  \langle S^\mathfrak{b}_\mathbf{R} \rangle &=& \left[ S - (\rho^\prime_\mathbf{R})^2 -(\rho^{\prime\prime}_\mathbf{R})^2 \right] \cos{\alpha} \sin{\mathbf{q}^\prime \mathbf{R}} \\
  &+& \sqrt{2S} \rho^\prime_\mathbf{R} \cos{\mathbf{q}^\prime \mathbf{R}} - \sqrt{2S} \rho^{\prime\prime}_\mathbf{R} \sin{\alpha} \sin{\mathbf{q}^\prime \mathbf{R}},   \nonumber \\
  \langle S^\mathfrak{c}_\mathbf{R} \rangle &=& \left[ S - (\rho^\prime_\mathbf{R})^2 -(\rho^{\prime\prime}_\mathbf{R})^2 \right] \sin{\alpha} \\ &+& \sqrt{2S} \rho^{\prime\prime}_\mathbf{R} \cos{\alpha} \nonumber
\end{eqnarray}
which have to be used in averaging over the disorder realizations (see, e.g., Ref.~\cite{utesov2015} for more details). The main results of the particular calculations are the following. There are magnetic Bragg peaks at momenta transfer $\bm{Q} = \pm (\mathbf{q} + \delta \mathbf{q}) + \bm{\tau} = \pm \mathbf{q}^\prime + \bm{\tau} $ and $ \bm{Q} = \bm{\tau}$, where $\bm{\tau}$ is a reciprocal lattice vector. In the conical phase, their spectral weights are proportional to $\cos^2{\alpha}$ and $\sin^2{\alpha}$, respectively. There are also small corrections proportional to $c$ to these spectral weights due to the disorder. Then, there is a diffuse scattering from defects. In the first order in $c$, it is related with the double Fourier transforms of $\overline{ \rho^\prime_{\mathbf{R}_1} \rho^\prime_{\mathbf{R}_2} }$, $\overline{ \rho^\prime_{\mathbf{R}_1} \rho^{\prime \prime}_{\mathbf{R}_2} } $, and $\overline{ \rho^{\prime \prime}_{\mathbf{R}_1} \rho^{\prime \prime}_{\mathbf{R}_2} }$ from a single defect, where the line denotes averaging over disorder realizations (see Ref.~\cite{utesov2015} for more details) and we also neglect terms containing products of more than two $\rho$. Due to the long-ranged character of the dipole field, the contribution from $\overline{ \rho^\prime_{\mathbf{R}_1} \rho^\prime_{\mathbf{R}_2} }$ is the most singular one. We obtain for it after tedious calculations
\begin{multline}
  \label{Diff1}
  \left(\frac{d \sigma}{d \Omega} \right)'
	\propto  N \frac{c}{3} S^2 \left( \frac{{\cal Q}_r}{J}\right)^2 \frac{1+\widehat{Q}_\mathfrak{c}^2}{2}\cos^2{\alpha} \sum_{\bm{\tau}} \sum_{\nu=x,y,z} {\mathfrak c}^2_\nu  \\
	\times \left(
	\frac{1-\cos{(Q_\nu+q^\prime_\nu-\tau_\nu)}}{\left({\bf Q}+\mathbf{q}^\prime-\bm{\tau}\right)^4}
	+
	\frac{1-\cos{(Q_\nu-q^\prime_\nu-\tau_\nu)}}{\left({\bf Q}-\mathbf{q}^\prime-\bm{\tau}\right)^4}
	\right),
\end{multline}
where ${\cal Q}_r$ is the ``charge'' given by Eq.~\eqref{Real5}. Thus, we conclude that the disorder leads to power-law singularities at Bragg peaks positions with $\bm{Q} = \pm \mathbf{q}^\prime + \bm{\tau} $. In contrast, there are only regular corrections from $\overline{ \rho^\prime_{\mathbf{R}_1} \rho^\prime_{\mathbf{R}_2} }$ to spectral weights of Bragg peaks at $ \bm{Q} = \bm{\tau}$. Our analysis shows that with the accuracy of our calculations $\overline{ \rho^\prime_{\mathbf{R}_1} \rho^{\prime \prime}_{\mathbf{R}_2} }$ gives zero. We derive the following correction to the cross-section from $\overline{ \rho^{\prime \prime}_{\mathbf{R}_1} \rho^{\prime \prime}_{\mathbf{R}_2} }$:
\begin{multline}
\label{Diff2}
   \left(\frac{d \sigma}{d \Omega} \right)''
	\propto N \frac{c}{3} S^2 \left( \frac{{\cal Q}_i}{J}\right)^2 \frac{1+\widehat{Q}_\mathfrak{c}^2}{2} \sin^4{\alpha} \cos^2{\alpha}  \\
	\times \sum_{\bm{\tau}} \sum_{\nu=x,y,z} {\mathfrak c}^4_\nu \Biggl(
	\frac{1+\cos{(Q_\nu+q^\prime_\nu-\tau_\nu)}}{\left[\left({\bf Q}+\mathbf{q}^\prime-\bm{\tau}\right)^2 + q^2 \cos^2{\alpha} \right]^2}  \\
	+
	\frac{1+\cos{(Q_\nu-q^\prime_\nu-\tau_\nu)}}{\left[\left({\bf Q}-\mathbf{q}^\prime-\bm{\tau}\right)^2 + q^2 \cos^2{\alpha} \right]^2}
	\Biggr)  \\
	+  N \frac{2 c}{3} S^2 \left( \frac{{\cal Q}_i}{J}\right)^2 (1-\widehat{Q}_\mathfrak{c}^2) \sin^2{\alpha} \cos^4{\alpha}  \\
	\times \sum_{\bm{\tau}} \sum_{\nu=x,y,z} {\mathfrak c}^4_\nu
	\frac{1+\cos{(Q_\nu-\tau_\nu)}}{\left[\left({\bf Q}-\bm{\tau}\right)^2 + q^2 \cos^2{\alpha} \right]^2},
\end{multline}
where the ``charge'' ${\cal Q}_i$ is given by Eq.~\eqref{Imag3}. Thus, all Bragg peaks acquire power-law decaying tails originating from the diffuse scattering cross section
\begin{equation}
\label{sigmares}
	\left(\frac{d \sigma}{d \Omega} \right)_{diff} = \left(\frac{d \sigma}{d \Omega} \right)'
	+
	\left(\frac{d \sigma}{d \Omega} \right)''.
\end{equation}

\section{Summary and conclusion}
\label{Conc}

In the present paper we theoretically discuss B20 helimagnets with defect bonds in external magnetic field $\bf H$ in the conical phase.
We assume that both exchange coupling and DMI are changed on defect bonds.
In pure system a conical spiral ordering appears with the spiral vector $\bf q$ directed along the field. We show that impurities lead to a distortion of the magnetic order
which can be represented at each site as variations of the spiral pitch and of the cone angle. In one-impurity problem, the distortion in the spiral pitch is long-ranged and it is governed by the Poisson equation for electric dipole whose solution is given by Eq.~\eqref{Real5}. The variation of the cone angle arises at finite magnetic field only and it is described by the screened Poisson equation for two electric charges, which solution has the form \eqref{Imag2}.

At finite defect concentration $c \ll 1$, in the first order in $c$ by averaging over disorder realizations we calculate corrections to the spiral vector and to the cone angle. We find that the spiral vector direction remains unchanged and the correction magnitude is independent on the magnetic field being given by Eq.~\eqref{Spiral1}. This lies in agreement with recent experimental findings of Refs.~\cite{grig2018,kinder2018} in Mn$_{1-x}$Fe$_x$Si at small $x$. The sign of the correction can be either positive, negative or zero, depending on the model parameters. The variation of the cone angle is given by Eq.~\eqref{Angle1}.

Cross section of the elastic neutron scattering is discussed. It is obtained that there are magnetic Bragg peaks at momenta transfer $\bm{Q} = \pm \mathbf{q}^\prime + \bm{\tau} $ and $ \bm{Q} = \bm{\tau}$, where $\mathbf{q}^\prime$ is renormalized spiral vector and $\bm{\tau}$ is a reciprocal lattice vector. It is shown that disorder leads to a diffuse elastic scattering having the form of power-law decaying tails centered at the Bragg peaks positions (see Eqs.~\eqref{Diff1}--\eqref{sigmares}). Unfortunately, the available experimental results on neutron scattering do not allow a detailed comparison with our findings. We hope that the latter can be useful in interpretation of further experiments.

Our results are inapplicable at strong magnetic fields, when the system is close to the transition to a glassy phase intervening between the ordered and the fully saturated states.



\begin{acknowledgments}

The reported study was funded by RFBR according to the research project 18-32-00083.

\end{acknowledgments}

\bibliography{bibliography}

\begin{thebibliography}{23}
\expandafter\ifx\csname natexlab\endcsname\relax\def\natexlab#1{#1}\fi
\expandafter\ifx\csname bibnamefont\endcsname\relax
  \def\bibnamefont#1{#1}\fi
\expandafter\ifx\csname bibfnamefont\endcsname\relax
  \def\bibfnamefont#1{#1}\fi
\expandafter\ifx\csname citenamefont\endcsname\relax
  \def\citenamefont#1{#1}\fi
\expandafter\ifx\csname url\endcsname\relax
  \def\url#1{\texttt{#1}}\fi
\expandafter\ifx\csname urlprefix\endcsname\relax\def\urlprefix{URL }\fi
\providecommand{\bibinfo}[2]{#2}
\providecommand{\eprint}[2][]{\url{#2}}

\bibitem[{\citenamefont{Dzyaloshinsky}(1958)}]{dzyal1958}
\bibinfo{author}{\bibfnamefont{I.}~\bibnamefont{Dzyaloshinsky}},
  \bibinfo{journal}{Journal of Physics and Chemistry of Solids}
  \textbf{\bibinfo{volume}{4}}, \bibinfo{pages}{241 } (\bibinfo{year}{1958}).

\bibitem[{\citenamefont{Moriya}(1960)}]{moriya1960}
\bibinfo{author}{\bibfnamefont{T.}~\bibnamefont{Moriya}},
  \bibinfo{journal}{Phys. Rev.} \textbf{\bibinfo{volume}{120}},
  \bibinfo{pages}{91} (\bibinfo{year}{1960}).

\bibitem[{\citenamefont{Dzyaloshinsky}(1964)}]{dzyal1964}
\bibinfo{author}{\bibfnamefont{I.}~\bibnamefont{Dzyaloshinsky}},
  \bibinfo{journal}{Zh. Eksp. Teor. Fiz.} \textbf{\bibinfo{volume}{19}},
  \bibinfo{pages}{960} (\bibinfo{year}{1964}).

\bibitem[{\citenamefont{Togawa et~al.}(2012)\citenamefont{Togawa, Koyama,
  Takayanagi, Mori, Kousaka, Akimitsu, Nishihara, Inoue, Ovchinnikov, and
  Kishine}}]{togawa}
\bibinfo{author}{\bibfnamefont{Y.}~\bibnamefont{Togawa}},
  \bibinfo{author}{\bibfnamefont{T.}~\bibnamefont{Koyama}},
  \bibinfo{author}{\bibfnamefont{K.}~\bibnamefont{Takayanagi}},
  \bibinfo{author}{\bibfnamefont{S.}~\bibnamefont{Mori}},
  \bibinfo{author}{\bibfnamefont{Y.}~\bibnamefont{Kousaka}},
  \bibinfo{author}{\bibfnamefont{J.}~\bibnamefont{Akimitsu}},
  \bibinfo{author}{\bibfnamefont{S.}~\bibnamefont{Nishihara}},
  \bibinfo{author}{\bibfnamefont{K.}~\bibnamefont{Inoue}},
  \bibinfo{author}{\bibfnamefont{A.~S.} \bibnamefont{Ovchinnikov}},
  \bibnamefont{and} \bibinfo{author}{\bibfnamefont{J.}~\bibnamefont{Kishine}},
  \bibinfo{journal}{Phys. Rev. Lett.} \textbf{\bibinfo{volume}{108}},
  \bibinfo{pages}{107202} (\bibinfo{year}{2012}).

\bibitem[{\citenamefont{M\"uhlbauer et~al.}(2009)\citenamefont{M\"uhlbauer,
  Binz, Jonietz, Pfleiderer, Rosch, Neubauer, Georgii, and B\"uni}}]{muhlbauer}
\bibinfo{author}{\bibfnamefont{S.}~\bibnamefont{M\"uhlbauer}},
  \bibinfo{author}{\bibfnamefont{B.}~\bibnamefont{Binz}},
  \bibinfo{author}{\bibfnamefont{F.}~\bibnamefont{Jonietz}},
  \bibinfo{author}{\bibfnamefont{C.}~\bibnamefont{Pfleiderer}},
  \bibinfo{author}{\bibfnamefont{A.}~\bibnamefont{Rosch}},
  \bibinfo{author}{\bibfnamefont{A.}~\bibnamefont{Neubauer}},
  \bibinfo{author}{\bibfnamefont{R.}~\bibnamefont{Georgii}}, \bibnamefont{and}
  \bibinfo{author}{\bibfnamefont{P.}~\bibnamefont{B\"uni}},
  \bibinfo{journal}{Science} \textbf{\bibinfo{volume}{323}},
  \bibinfo{pages}{915} (\bibinfo{year}{2009}).

\bibitem[{\citenamefont{Bauer et~al.}(2010)\citenamefont{Bauer, Neubauer,
  Franz, M\"unzer, Garst, and Pfleiderer}}]{bauer2010}
\bibinfo{author}{\bibfnamefont{A.}~\bibnamefont{Bauer}},
  \bibinfo{author}{\bibfnamefont{A.}~\bibnamefont{Neubauer}},
  \bibinfo{author}{\bibfnamefont{C.}~\bibnamefont{Franz}},
  \bibinfo{author}{\bibfnamefont{W.}~\bibnamefont{M\"unzer}},
  \bibinfo{author}{\bibfnamefont{M.}~\bibnamefont{Garst}}, \bibnamefont{and}
  \bibinfo{author}{\bibfnamefont{C.}~\bibnamefont{Pfleiderer}},
  \bibinfo{journal}{Phys. Rev. B} \textbf{\bibinfo{volume}{82}},
  \bibinfo{pages}{064404} (\bibinfo{year}{2010}),
  \urlprefix\url{https://link.aps.org/doi/10.1103/PhysRevB.82.064404}.

\bibitem[{\citenamefont{Glushkov et~al.}(2015)\citenamefont{Glushkov, Lobanova,
  Ivanov, Voronov, Dyadkin, Chubova, Grigoriev, and Demishev}}]{glushkov2015}
\bibinfo{author}{\bibfnamefont{V.~V.} \bibnamefont{Glushkov}},
  \bibinfo{author}{\bibfnamefont{I.~I.} \bibnamefont{Lobanova}},
  \bibinfo{author}{\bibfnamefont{V.~Y.} \bibnamefont{Ivanov}},
  \bibinfo{author}{\bibfnamefont{V.~V.} \bibnamefont{Voronov}},
  \bibinfo{author}{\bibfnamefont{V.~A.} \bibnamefont{Dyadkin}},
  \bibinfo{author}{\bibfnamefont{N.~M.} \bibnamefont{Chubova}},
  \bibinfo{author}{\bibfnamefont{S.~V.} \bibnamefont{Grigoriev}},
  \bibnamefont{and} \bibinfo{author}{\bibfnamefont{S.~V.}
  \bibnamefont{Demishev}}, \bibinfo{journal}{Phys. Rev. Lett.}
  \textbf{\bibinfo{volume}{115}}, \bibinfo{pages}{256601}
  (\bibinfo{year}{2015}),
  \urlprefix\url{https://link.aps.org/doi/10.1103/PhysRevLett.115.256601}.

\bibitem[{\citenamefont{Grigoriev et~al.}(2013)\citenamefont{Grigoriev,
  Potapova, Siegfried, Dyadkin, Moskvin, Dmitriev, Menzel, Dewhurst,
  Chernyshov, Sadykov et~al.}}]{grig2013}
\bibinfo{author}{\bibfnamefont{S.~V.} \bibnamefont{Grigoriev}},
  \bibinfo{author}{\bibfnamefont{N.~M.} \bibnamefont{Potapova}},
  \bibinfo{author}{\bibfnamefont{S.-A.} \bibnamefont{Siegfried}},
  \bibinfo{author}{\bibfnamefont{V.~A.} \bibnamefont{Dyadkin}},
  \bibinfo{author}{\bibfnamefont{E.~V.} \bibnamefont{Moskvin}},
  \bibinfo{author}{\bibfnamefont{V.}~\bibnamefont{Dmitriev}},
  \bibinfo{author}{\bibfnamefont{D.}~\bibnamefont{Menzel}},
  \bibinfo{author}{\bibfnamefont{C.~D.} \bibnamefont{Dewhurst}},
  \bibinfo{author}{\bibfnamefont{D.}~\bibnamefont{Chernyshov}},
  \bibinfo{author}{\bibfnamefont{R.~A.} \bibnamefont{Sadykov}},
  \bibnamefont{et~al.}, \bibinfo{journal}{Phys. Rev. Lett.}
  \textbf{\bibinfo{volume}{110}}, \bibinfo{pages}{207201}
  (\bibinfo{year}{2013}).

\bibitem[{\citenamefont{Grigoriev et~al.}(2015)\citenamefont{Grigoriev,
  Sukhanov, and Maleyev}}]{grig2015}
\bibinfo{author}{\bibfnamefont{S.~V.} \bibnamefont{Grigoriev}},
  \bibinfo{author}{\bibfnamefont{A.~S.} \bibnamefont{Sukhanov}},
  \bibnamefont{and} \bibinfo{author}{\bibfnamefont{S.~V.}
  \bibnamefont{Maleyev}}, \bibinfo{journal}{Phys. Rev. B}
  \textbf{\bibinfo{volume}{91}}, \bibinfo{pages}{224429}
  (\bibinfo{year}{2015}),
  \urlprefix\url{https://link.aps.org/doi/10.1103/PhysRevB.91.224429}.

\bibitem[{\citenamefont{Kikuchi et~al.}(2016)\citenamefont{Kikuchi, Koretsune,
  Arita, and Tatara}}]{kikuchi2016}
\bibinfo{author}{\bibfnamefont{T.}~\bibnamefont{Kikuchi}},
  \bibinfo{author}{\bibfnamefont{T.}~\bibnamefont{Koretsune}},
  \bibinfo{author}{\bibfnamefont{R.}~\bibnamefont{Arita}}, \bibnamefont{and}
  \bibinfo{author}{\bibfnamefont{G.}~\bibnamefont{Tatara}},
  \bibinfo{journal}{Phys. Rev. Lett.} \textbf{\bibinfo{volume}{116}},
  \bibinfo{pages}{247201} (\bibinfo{year}{2016}),
  \urlprefix\url{https://link.aps.org/doi/10.1103/PhysRevLett.116.247201}.

\bibitem[{\citenamefont{Utesov et~al.}(2015)\citenamefont{Utesov, Sizanov, and
  Syromyatnikov}}]{utesov2015}
\bibinfo{author}{\bibfnamefont{O.~I.} \bibnamefont{Utesov}},
  \bibinfo{author}{\bibfnamefont{A.~V.} \bibnamefont{Sizanov}},
  \bibnamefont{and} \bibinfo{author}{\bibfnamefont{A.~V.}
  \bibnamefont{Syromyatnikov}}, \bibinfo{journal}{Phys. Rev. B}
  \textbf{\bibinfo{volume}{92}}, \bibinfo{pages}{125110}
  (\bibinfo{year}{2015}),
  \urlprefix\url{https://link.aps.org/doi/10.1103/PhysRevB.92.125110}.

\bibitem[{\citenamefont{Grigoriev et~al.}(2018)\citenamefont{Grigoriev,
  Altynbaev, Siegfried, Pschenichnyi, Menzel, Heinemann, and
  Chaboussant}}]{grig2018}
\bibinfo{author}{\bibfnamefont{S.~V.} \bibnamefont{Grigoriev}},
  \bibinfo{author}{\bibfnamefont{E.~V.} \bibnamefont{Altynbaev}},
  \bibinfo{author}{\bibfnamefont{S.-A.} \bibnamefont{Siegfried}},
  \bibinfo{author}{\bibfnamefont{K.~A.} \bibnamefont{Pschenichnyi}},
  \bibinfo{author}{\bibfnamefont{D.}~\bibnamefont{Menzel}},
  \bibinfo{author}{\bibfnamefont{A.}~\bibnamefont{Heinemann}},
  \bibnamefont{and}
  \bibinfo{author}{\bibfnamefont{G.}~\bibnamefont{Chaboussant}},
  \bibinfo{journal}{Phys. Rev. B} \textbf{\bibinfo{volume}{97}},
  \bibinfo{pages}{024409} (\bibinfo{year}{2018}),
  \urlprefix\url{https://link.aps.org/doi/10.1103/PhysRevB.97.024409}.

\bibitem[{\citenamefont{Kindervater et~al.}(2018)\citenamefont{Kindervater,
  Adams, Bauer, Haslbeck, Chacon, Muhlbauer, Jonietz, Neubauer, Gasser, Nagy
  et~al.}}]{kinder2018}
\bibinfo{author}{\bibfnamefont{J.}~\bibnamefont{Kindervater}},
  \bibinfo{author}{\bibfnamefont{T.}~\bibnamefont{Adams}},
  \bibinfo{author}{\bibfnamefont{A.}~\bibnamefont{Bauer}},
  \bibinfo{author}{\bibfnamefont{F.}~\bibnamefont{Haslbeck}},
  \bibinfo{author}{\bibfnamefont{A.}~\bibnamefont{Chacon}},
  \bibinfo{author}{\bibfnamefont{S.}~\bibnamefont{Muhlbauer}},
  \bibinfo{author}{\bibfnamefont{F.}~\bibnamefont{Jonietz}},
  \bibinfo{author}{\bibfnamefont{A.}~\bibnamefont{Neubauer}},
  \bibinfo{author}{\bibfnamefont{U.}~\bibnamefont{Gasser}},
  \bibinfo{author}{\bibfnamefont{G.}~\bibnamefont{Nagy}}, \bibnamefont{et~al.},
  \emph{\bibinfo{title}{Evolution of magnetocrystalline anisotropies in
  mn$_{1-x}$fe$_x$si and mn$_{1-x}$co$_x$si as observed in small-angle neutron
  scattering}} (\bibinfo{year}{2018}), \eprint{arXiv:1811.12379}.

\bibitem[{\citenamefont{Bak and Jensen}(1980)}]{Bak}
\bibinfo{author}{\bibfnamefont{P.}~\bibnamefont{Bak}} \bibnamefont{and}
  \bibinfo{author}{\bibfnamefont{M.~H.} \bibnamefont{Jensen}},
  \bibinfo{journal}{Journal of Physics C: Solid State Physics}
  \textbf{\bibinfo{volume}{13}}, \bibinfo{pages}{L881} (\bibinfo{year}{1980}).

\bibitem[{\citenamefont{Yi et~al.}(2009)\citenamefont{Yi, Onoda, Nagaosa, and
  Han}}]{yi2009}
\bibinfo{author}{\bibfnamefont{S.~D.} \bibnamefont{Yi}},
  \bibinfo{author}{\bibfnamefont{S.}~\bibnamefont{Onoda}},
  \bibinfo{author}{\bibfnamefont{N.}~\bibnamefont{Nagaosa}}, \bibnamefont{and}
  \bibinfo{author}{\bibfnamefont{J.~H.} \bibnamefont{Han}},
  \bibinfo{journal}{Phys. Rev. B} \textbf{\bibinfo{volume}{80}},
  \bibinfo{pages}{054416} (\bibinfo{year}{2009}),
  \urlprefix\url{https://link.aps.org/doi/10.1103/PhysRevB.80.054416}.

\bibitem[{\citenamefont{Nakanishi et~al.}(1980)\citenamefont{Nakanishi, Yanase,
  Hasegawa, and Kataoka}}]{nakanishi1980}
\bibinfo{author}{\bibfnamefont{O.}~\bibnamefont{Nakanishi}},
  \bibinfo{author}{\bibfnamefont{A.}~\bibnamefont{Yanase}},
  \bibinfo{author}{\bibfnamefont{A.}~\bibnamefont{Hasegawa}}, \bibnamefont{and}
  \bibinfo{author}{\bibfnamefont{M.}~\bibnamefont{Kataoka}},
  \bibinfo{journal}{Solid State Communications} \textbf{\bibinfo{volume}{35}},
  \bibinfo{pages}{995 } (\bibinfo{year}{1980}), ISSN \bibinfo{issn}{0038-1098},
  \urlprefix\url{http://www.sciencedirect.com/science/article/pii/0038109880910042}.

\bibitem[{\citenamefont{Kaplan}(1961)}]{kaplan}
\bibinfo{author}{\bibfnamefont{T.~A.} \bibnamefont{Kaplan}},
  \bibinfo{journal}{Phys. Rev.} \textbf{\bibinfo{volume}{124}},
  \bibinfo{pages}{329} (\bibinfo{year}{1961}),
  \urlprefix\url{https://link.aps.org/doi/10.1103/PhysRev.124.329}.

\bibitem[{\citenamefont{Maleyev}(2006)}]{maleyev}
\bibinfo{author}{\bibfnamefont{S.~V.} \bibnamefont{Maleyev}},
  \bibinfo{journal}{Phys. Rev. B} \textbf{\bibinfo{volume}{73}},
  \bibinfo{pages}{174402} (\bibinfo{year}{2006}).

\bibitem[{\citenamefont{Holstein and Primakoff}(1940)}]{holstein1940}
\bibinfo{author}{\bibfnamefont{T.}~\bibnamefont{Holstein}} \bibnamefont{and}
  \bibinfo{author}{\bibfnamefont{H.}~\bibnamefont{Primakoff}},
  \bibinfo{journal}{Phys. Rev.} \textbf{\bibinfo{volume}{58}},
  \bibinfo{pages}{1098} (\bibinfo{year}{1940}),
  \urlprefix\url{https://link.aps.org/doi/10.1103/PhysRev.58.1098}.

\bibitem[{\citenamefont{Pollet et~al.}(2009)\citenamefont{Pollet, Prokof'ev,
  Svistunov, and Troyer}}]{theorem}
\bibinfo{author}{\bibfnamefont{L.}~\bibnamefont{Pollet}},
  \bibinfo{author}{\bibfnamefont{N.~V.} \bibnamefont{Prokof'ev}},
  \bibinfo{author}{\bibfnamefont{B.~V.} \bibnamefont{Svistunov}},
  \bibnamefont{and} \bibinfo{author}{\bibfnamefont{M.}~\bibnamefont{Troyer}},
  \bibinfo{journal}{Phys. Rev. Lett.} \textbf{\bibinfo{volume}{103}},
  \bibinfo{pages}{140402} (\bibinfo{year}{2009}).

\bibitem[{\citenamefont{Syromyatnikov and Sizanov}(2017)}]{syromyat}
\bibinfo{author}{\bibfnamefont{A.~V.} \bibnamefont{Syromyatnikov}}
  \bibnamefont{and} \bibinfo{author}{\bibfnamefont{A.~V.}
  \bibnamefont{Sizanov}}, \bibinfo{journal}{Phys. Rev. B}
  \textbf{\bibinfo{volume}{95}}, \bibinfo{pages}{014206}
  (\bibinfo{year}{2017}), \bibinfo{note}{and references therein},
  \urlprefix\url{https://link.aps.org/doi/10.1103/PhysRevB.95.014206}.

\bibitem[{\citenamefont{Lowesey}(1987)}]{Lowesey}
\bibinfo{author}{\bibfnamefont{S.~W.} \bibnamefont{Lowesey}},
  \emph{\bibinfo{title}{Theory of Neutron Scattering by Condensed Matter}}
  (\bibinfo{publisher}{Oxford University Press, Oxford}, \bibinfo{year}{1987}).

\bibitem[{\citenamefont{Demishev et~al.}(2011)\citenamefont{Demishev, Semeno,
  Bogach, Glushkov, Sluchanko, Samarin, and Chernobrovkin}}]{demishev2011}
\bibinfo{author}{\bibfnamefont{S.~V.} \bibnamefont{Demishev}},
  \bibinfo{author}{\bibfnamefont{A.~V.} \bibnamefont{Semeno}},
  \bibinfo{author}{\bibfnamefont{A.~V.} \bibnamefont{Bogach}},
  \bibinfo{author}{\bibfnamefont{V.~V.} \bibnamefont{Glushkov}},
  \bibinfo{author}{\bibfnamefont{N.~E.} \bibnamefont{Sluchanko}},
  \bibinfo{author}{\bibfnamefont{N.~A.} \bibnamefont{Samarin}},
  \bibnamefont{and} \bibinfo{author}{\bibfnamefont{A.~L.}
  \bibnamefont{Chernobrovkin}}, \bibinfo{journal}{JETP Letters}
  \textbf{\bibinfo{volume}{93}}, \bibinfo{pages}{213} (\bibinfo{year}{2011}),
  ISSN \bibinfo{issn}{1090-6487},
  \urlprefix\url{https://doi.org/10.1134/S0021364011040072}.

\end{thebibliography}

\end{document}